\documentclass[aps,preprint,superscriptaddress,groupedaddress]{revtex4}  

\usepackage{graphicx}  
\usepackage{dcolumn}   
\usepackage{bm}        
\usepackage{amssymb}   
\usepackage{amsmath}
\usepackage{color}
\usepackage{dsfont}

\newcommand{\be}{\begin{equation}}
\newcommand{\ee}{\end{equation}}
\newcommand{\bea}{\begin{eqnarray}}
\newcommand{\eea}{\end{eqnarray}}

\newcommand{\nnu}{\nonumber\\}

\newcommand{\os}{\overline\Sigma}
\newcommand{\s}{\Sigma}
\newcommand{\Sigb}{{\overline\Sigma}}
\newcommand{\oot}{\overline {126}}

\newcommand{\ovl}{\overline}

\begin{document}

\title{ New Minimal SO(10) GUT :
A Theory for All Epochs\footnote{To appear in the proceedings of
UNICOS2014  \cite{unicosproc}}}
\author{ Charanjit S. Aulakh\footnote{aulakh@iisermohali.ac.in;
14.139.227.202/Faculty/aulakh/}}
\affiliation{Dept. of Physics, Panjab University\\
 Chandigarh, 160014, India}
\affiliation{Indian Institute of Science Education and Research
Mohali,\\ Sector 81, S. A. S. Nagar, Manauli PO 140306, India }

\begin{abstract}
The Supersymmetric SO(10) theory (``NMSO(10)GUT")   based on
the\hfil\break  ${\bf{210+126 +\oot}}$ Higgs system proposed in
1982 has evolved into a realistic theory capable of fitting the
known low energy Particle Physics data besides providing a Dark
matter candidate and embedding Inflationary Cosmology. It
dynamically resolves longstanding issues such as fast dimension
five operator mediated proton decay in Susy GUTs by allowing
explicit and complete calculation of crucial threshold effects at
$M_{Susy}$ and $M_{GUT}$ in terms of fundamental parameters. This
shows that SO(10)  Yukawas responsible for observed fermion masses
as well as operator dimension 5 mediated proton decay  can be
highly suppressed on a ``Higgs dissolution edge'' in the parameter
space of GUTs with rich superheavy spectra. This novel and
generically relevant  result highlights
  the need for every realistic UV completion model
with a large/infinite number of heavy fields coupled to the light
Higgs doublets to explicitly account for the large wave function
renormalization effects on emergent light Higgs fields in order to
be considered a quantitatively well defined candidate UV
completion. The NMSGUT predicts   large  soft Susy breaking
trilinear couplings and distinctive sparticle spectra. Measurable
or near measurable level of tensor perturbations- and thus large
Inflaton mass scale-  may be accommodated by Supersymetric Seesaw
inflation within the NMSGUT based on an LHN flat direction
Inflaton if the Higgs component contains contributions from heavy
Higgs components. Successful NMSGUT fits suggest a
\emph{renormalizable} Yukawon Ultra minimal gauged theory of
flavor based upon the NMSGUT Higgs structure.
\end{abstract}



\maketitle


\section{Introduction}
 Grand Unification theories (GUTs)  have  seen some 40 summers
 since the basic idea was first  proposed by Pati and Salam in
 1974\cite{patisalam} yet retain
  their  attraction   as the most obvious progression of the gauge
   logic embodied in   the standard model. Indeed, within Particle physics,  the only  tangible
  and  clear  hints of Physics beyond the
  Standard model are the remarkable convergence of  gauge couplings
  in the   Unification regime   :  $ 10^{15} - 10^{17} $ GeV
  and the existence of neutrino masses
 (initially thought to be  superfluous in both the
 Electro-weak  and   GUT contexts). To this one should perhaps
 append the convergence of the pivotal\cite{realcore}  third
  generation Yukawa couplings in the MSSM at high $\tan\beta$.
   The   same  gauge logic  that structures the SM (and justifies the zeroth order decoupling
 of  right handed neutrinos) also makes the
 Type I  seesaw mechanism\cite{seesaw}  the most natural rationale for the
  milli-eV range neutrino masses actually observed.
  Supersymmetry seems essential to the structural
  stability of the SM and experience\cite{MSLRMs,rparso10} with supersymmetric LR models
  and  renormalizable  SO(10) GUTs shows it   ensures
  complete calculability for   UV completions of the MSSM.
     Thus non-discovery of sparticles in the searches so far
  has still not dimmed the ardor of its many adherents.
  Arguments for considering some variety of the MSSM as the
  effective theory of any GUT became all the more cogent when, with the observation of
  solar and atmospheric neutrino oscillations at Super-Kamiokande,  the
  scale of B-L breaking associated with a renormalizable seesaw
  mechanism emerged     ($V_{EW}^2/(10^{-12} GeV)\sim 10^{14}$ GeV) in the GUT  ball-park.
   The convergence of these two compelling arguments thus
  hints at a deep interconnection between B and L violation in
  a Supersymmetric   Grand Unified framework\cite{sunimnu,millenium}.  It is SO(10)
  \emph{not} SU(5) GUTs that provide the most natural GUT
  framework, free of  gauge  singlets (which are really opaque, and thus anathemic,
   to the gauge logic pursued so successfully in the SM),  for
   Type I (and  also Type II ) seesaw. Thus eventually  SO(10) GUTs
   displaced SU(5) theories and achieved a long delayed vindication
    and recognition as minimal UV complete  frameworks
      for  neutrino oscillations. It was only natural
  that our pursuit of consistent  minimal Left right supersymmetric
  R-parity preserving models\cite{MSLRMs} from the mid-nineties led in short
  order\cite{rparso10}   after the epochal measurements of Super-Kamiokande
  to a   realization  that the very first complete supersymmetric SO(10)
  GUT\cite{aulmoh,ckn} proposed way back in 1982, just after the
  Georgi-Dimopoulos minimal Susy SU(5) model\cite{georgidimo}, was in fact the
  Minimal Supersymmetric GUT(MSGUT)\cite{abmsv}.
 This model is now called\cite{nmsgut}- due to a transient  glory of a truncated
 version with only $\bf{10+{\overline{126}}}$ \emph{but not
 }${\bf{120}}$ coupling to matter fermion bilinears as the Minimal
 theory\cite{BabMoh} -    the New/Next MSGUT(NMSGUT). In the best Popperian
 mode it   has survived another decade of detailed investigation
 of its ability to fit all available
  SM, gauge,  fermion and neutrino data,  as well as the consistency of
 one-loop threshold corrections at both low and high energy scales made possible
  by its extreme parameter economy(a.k.a. minimality)  and full calculability : simple
  virtues that are, alas,
  all too rare among the plethora of its (non-minimal) competitors.
  The model is based  on a ${\bf{210\oplus 126\oplus {\overline{126}}}}$ GUT Higgs
 system and already in 1982\cite{aulmoh}  showed clearly that   prevention of a RG flow catastrophe due to   large
 pseudo-goldstone supermultiplets  requires a single step breaking of Susy SO(10)
 to the MSSM  :   later rediscovered in the context of another
 R-parity preserving but   non-minimal model\cite{rparso10}. Being
 a complete theory of gauge physics the model is capable of
 fitting and/or  predicting most, if not all, of the commonly considered varieties
 of BSM physics, including  neutrino masses,  the g-2 muon anomaly, LSP dark
 matter, B violation (at acceptable rates : see below),
 Leptogenesis driven by the parameters controlling neutrino mass
 etc. Remarkably it also comes close to providing a workable
 embedding of inflation based upon the  D-flat directions involving Lepton and Higgs doublet
 fields  naturally present and contributing to the effective MSSM
 below the GUT scale.  The model also predicts a distinctive and
 characteristic \emph{normal} hierarchy of sfermion generations
 discoverable by the LHC or its successors. As striking was its
 early indication, on the basis of the parameters required for a
 successful fermion fit, that the   trilinear soft susy breaking parameter $A_0$
   \emph{must } be large (as far back as  2008\cite{nmsgutarxiv08} i.e
 well before Higgs mass discovery made such a large $A_0$  respectable
 rather than ludicrously fine tuned: as then argued by naturalists).

The fit the of  low energy data in terms of fundamental parameters
obtained in  the  NMSGUT  must further be consistent with, indeed
predict, a vast range of phenomena ranging from exotic processes
 such as Baryon number  and Lepton flavour violation to
 cosmological Leptogenesis, Dark Matter relic density and
 inflation. The demonstration of the feasibility of  such comprehensive
 fits of the SM data directly in terms of Susy GUT  parameters
 marks the NMSGUT    as being cast in the mould of the SM
 rather than its piece meal generalizations.
  The  most striking themes and results that have emerged recently from
 our detailed  investigation of the  structural freedoms-in-necessity
granted by the full  calculability  NMSGUT are:

 \begin{enumerate}
 \item A simple generic mechanism  for suppression of the long
 problematic\cite{dim5ops} $d=5$ super fast proton decay generic to Susy GUTs.

 \item An embedding of  high scale supersymmetric renormalizable
   inflexion\cite{ssI} based upon the
 left Lepton(L)   -Higgs doublet(H) - conjugate Neutrino($N\equiv \nu^c_L$)
 flat direction (labelled by the LHN chiral invariant) and  utilizing
 the involvement of heavy partners  of the MSSM Higgs. This may prove very
 attractive in the context of the BICEP2 driven  focus on high
 scale inflation.

 \item Completely novel ``Grand Yukawonification'' models based
 upon gauging of an O(3) subgroup of the  U(3) flavour symmetry of
 the SO(10) GUT fermion kinetic  terms.    This flavour symmetry is
 naturally broken at the GUT scale by the NMSGUT Higgs fields which
  promote themselves rather naturally to also carry the function of being Yukawons
   i.e. fields whose vevs determine the Yukawa couplings in
   the effective MSSM besides simultaneously  breaking SO(10) \emph{while maintaining
   renormalizability}.

 \item The models in item 3. above are made possible by a novel
 conflation of  supergravity  mediated  supersymmetry breaking
 with the breaking of gauged  flavour symmetry in the Hidden sector of the
 model using the so called Bajc-Melfo calculable metastable
 Susy breaking vacua\cite{gyuk,bmvacua}.

\end{enumerate}

Thus the NMSGUT potentially provides a completely realistic and
predictive theory of  particle physics in all energy ranges and
cosmological epochs. We have even speculated\cite{trmin} that the
Landau pole in the NMSGUT gauge coupling which occurs quite near
the Planck scale should  be interpreted as a physical cutoff on
the perturbative dynamics  associated with the scale where the
NMSGUT  condenses(via a strong coupled supersymmetric dynamics) in
the ultra violet. This scale could then function as the Planck
scale of an effective  induced supergravity  at large length
scales emergent from  the NMSGUT when the metric and gravitino
fields introduced to define a coordinate independent microscopic
GUT acquire kinetic terms due to quantum effects. Since items 2)
3) are published as \cite{ssI,gyuk} and are also  reported in
these proceedings in the contributions of my collaborators Ila
Garg and Charanjit Kaur respectively, I will touch upon them only
briefly but focus on  the results in 1), 4) and discuss some of
their implications while referring the reader to the published
papers\cite{abmsv,ag1,ag2,blmdm,bmsv,nmsgut} for details.

\section{NMSGUT basics}

 The NMSGUT superpotential  is built from the quadratic
 SO(10) invariants and associated mass parameters
 \bea
 m: {\bf{210}}^{\bf{2}} \quad ;\quad M : {\bf{126\cdot{\overline {126}}}}
\quad ;\quad M_H : {\bf{10}}^{\bf{2}}\quad;\quad m_{\Theta}
:{\bf{120}}^{\bf{2}} \eea

and trilinear couplings  :
 \bea
 \lambda &:& {\bf{210}}^{\bf{3}} \qquad ; \qquad \eta :
 {\bf{210\cdot 126\cdot{\overline {126}}}}
 \qquad;\qquad \rho :{\bf{120\cdot 120 \cdot{ { 210}}}}
\nnu k &:& {\bf{ 10\cdot 120\cdot{ {210}}}} \qquad;\qquad \gamma
\oplus {\bar\gamma} : {\bf{10 \cdot 210}\cdot(126 \oplus{\overline
{126}}}) \nnu \zeta \oplus {\bar\zeta} &:& {\bf{120 \cdot
210}\cdot(126 \oplus {\overline {126}}})\nnu && {\bf{{16}_{A}.
{16}_{B}}}.(h_{AB} \mathbf{10} + f_{AB}\mathbf{{\overline {126}}}
+g_{AB} \mathbf{120} )
 \eea
 The couplings $h,f(g)$ are complex (anti)-symmetric in the flavour
 indices due to properties of the SO(10) Clifford algebra.
  Either  $h$ or $f$  is chosen
  real and diagonal  using the $U(3)$ flavor symmetry
   of the matter kinetic terms. The
 matter  Yukawas thus contain 21 real
parameters. Five   phases  say  of
 $m,M, \lambda ,\gamma,\bar\gamma$ are set to zero by phase conventions.
 One (complex parameter)   say
 $M_H$, is  fine tuned to keep two  Higgs doublets
 ($H,{\overline{H}}$) of the effective MSSM light,
  leaving  23 magnitudes and
15 phases as parameters. Fine tuning fixes  $H,{\overline{H}}$
composition as a mixture  of the (6 pairs of the MSSM type)
 doublet fields in the GUT as a function of the superpotential
 parameters entering the null vectors of the
  $G_{3,2,1}$ irreps $[1,2,\pm 1]$ mass terms. The mixture is
  described by the so called `` Higgs fractions''
  \cite{abmsv,ag2}.

The GUT scale vevs( units  $m/\lambda$)  are known  functions of
$x$  which is a  solution   of the cubic ($\xi ={{ \lambda M}\over
{\eta m}} $) \be 8 x^3 - 15 x^2 + 14 x -3 +\xi (1-x)^2=0
\label{cubic} \ee

The complete set of GUT scale mass matrices for the 26 different
MSSM irrep types and tree level low energy effective
Superpotential was calculated \cite{nmsgut} extending the result
for the MSGUT\cite{ag1,ag2,bmsv,fuku}. Using these and two loop RG
flows the  gauge, superoptential  and soft susy breaking
parameters at GUT scales can be matched, using a downhill simplex
search procedure, to the known values of the MSSM data described
above\cite{nmsgut}. This also yields a mini-split (10-100 TeV)
supersymmetry sparticle spectrum with large $A_0,\mu$ parameters,
light gauginos and Bino LSP, super heavy Higgsinos, sfermions in
tens of TeV and sometimes a smuon(or other sfermion)  light enough
(i.e within 10\% of the light  Bino LSP)  to co-annihilate with it
and provide acceptable dark matter relic density\cite{ilaPhD}. The
light smuon case is obviously attractive if Supersymmetry is
called upon to explain the muon magnetic moment anomaly and
emerges in the right ball park for these solutions. Flavour
violation in the Quark and Lepton sectors is also well controlled
because of the multi-TeV masses of most of the sfermions. Moreover
$A_0$ is required to be large to allow the $b$ quark mass to be
fitted and this was found\cite{nmsgut,nmsgutarxiv08} 
serendipitously \emph{before} Higgs discovery made it a
requirement for Supersymmetry.

\section{GUT scale threshold corrections and Baryon decay rate}

The SO(10) parameters determined by the realistic fit are
substituted into the effective (dimension four)  superpotential
describing Baryon violation which,  after RG flow to low energies,
is  used to calculate the proton decay rate. As is well known the
generic result gives a lifetime of some $10^{27}$ years i.e some 7
orders of magnitude shorter than the current limits from the Super
Kamiokande experiment. The beautifully complete and predictive
fits are thus  of little use if this problem remains
unsurmountable. However, as in  previous tight spots,  the NMSGUT
points out a convincing and illuminating, generically applicable
and dynamical, pathway out of the difficulty : precisely because
of the available explicit solution described above.

Superpotential parameters   renormalize  only by  wave function
corrections. In the computable  basis sets where heavy
supermultiplet masses are diagonal a generic heavy field type
$\Phi$ (conjugate $\overline \Phi$)  mass matrix  diagonalizes as
:
 \bea{\overline{\Phi }}= U^{\Phi}{\overline{\Phi' }} \quad ;
 \qquad {\Phi } = V^{\Phi}\Phi'\quad \Rightarrow \quad
 {\overline{\Phi}}^T M \Phi ={\overline{\Phi'}}^T M_{Diag}
 \Phi'\eea

Circulation of heavy  within light field propagators entering the
matter($f_A,f^c_A$)-Higgs  Yukawa vertices : ${\cal L} = [f_c^T
Y_f f H_f]_F + H.c. + $ implies \cite{wright} a finite wave
function renormalization in the fermion and Higgs Kinetic terms
\bea {\cal L}=[\sum_{A,B}( {\bar f}_A^\dagger (Z_{\bar f})_A^B
{\bar f}_B +{f}_A^\dagger (Z_{f})_A^B {f}_B ) + H^\dagger Z_H H +
{\ovl H}^\dagger Z_{\ovl H} {\ovl H}]_D +..\eea

Unitary matrices $U_{Z_f},{\ovl U}_{Z_{\bar f}}$   diagonalize
($U^\dagger Z U= \Lambda_Z$) $Z_{f,\bar f}$ to positive definite
form $\Lambda_{Z_f,Z_{\bar f}}$. We define a new basis to put the
Kinetic terms of the light matter and Higgs fields in canonical
form :\bea f &=& U_{Z_f} \Lambda^{-\frac{1}{2}}_{Z_f} \tilde f
=\tilde U_{Z_f} \tilde f\qquad ; \qquad \bar f= U_{Z_{\bar f}}
\Lambda^{-\frac{1}{2}}_{Z_{\bar f}} \tilde{\bar f}=\tilde
U_{Z_{\bar {f}}} \tilde {\bar f}\nnu H&=&\frac{\widetilde
H}{\sqrt{Z_H}}\qquad ; \qquad \ovl H=\frac{\widetilde {\ovl
H}}{\sqrt{Z_{\ovl H}}}\label{f2ftilde}\eea Thus when matching to
the effective MSSM it is  Yukawa couplings of the effective MSSM
$\tilde Y_f$:  \be\tilde Y_f= \Lambda_{Z_{\bar f}}^{-\frac{1}{2}}
U_{Z_{\bar f}}^T {\frac{Y_f}{\sqrt{Z_{H_f}}}} U_{Z_f}
\Lambda_{Z_f}^{-\frac{1}{2}} = \tilde{U}_{Z_{\bar f}}^T
{\frac{Y_f}{\sqrt{Z_{H_f}}}} \tilde{U}_{Z_f} \label{Ynutilde}\ee
 and not the original tree level ones that match the MSSM at the matching scale.

Light Chiral field $\Phi_i$ the corrections have generic form ($Z=
1 -{\cal K}$) :\bea{\cal K}_i^j=- {\frac{g_{10}^2}{8 \pi^2}}
\sum_\alpha {Q^\alpha_{ik}}^* {Q^\alpha_{kj}} F(m_\alpha,m_k)
+{\frac{1}{32 \pi^2}}\sum_{kl} Y_{ikl} Y_{jkl}^* F(m_k,m_l) \eea

where ${\cal L}= g_{10} \, Q^\alpha_{ik} \psi^\dagger_i
{\gamma^\mu A_\mu}^\alpha \psi_k$ describes the generic gauge
coupling  and $W={\frac{1}{6}} Y_{ijk}\Phi_i \Phi_j \Phi_k$ the
generic Yukawa couplings, while   $F(m_1,m_2)$  are   symmetric
1-loop  Passarino-Veltman functions.

Crucially  the SO(10) Yukawa couplings $(h,f,g)_{AB}$ also enter
into the coefficients $L_{ABCD},R_{ABCD}$ of the $d=5 $ baryon
decay operators in the effective superpotential  obtained by
integrating out the heavy chiral supermultiplets that mediate
baryon decay\cite{ag1,ag2,nmsgut}.

   $\tilde Y_f$ must be diagonalized to mass
basis (denoted by primes) so that  $d=5, \Delta B =\pm 1$ decay
operator coefficients become \bea L_{ABCD}' &&=\sum_{a,b,c,d}
L_{abcd} (\tilde U_Q')_{aA} (\tilde U_Q')_{bB}(\tilde U_Q')_{cC}
(\tilde U_L')_{dD} \nnu R_{ABCD}' &&=\sum_{a,b,c,d} R_{abcd}
(\tilde U_{\bar e}')_{aA} (\tilde U_{\bar u}')_{bB}(\tilde U_{\bar
u}')_{cC} (\tilde U_{\bar d}')_{dD} \eea

The search for a fit with the  constraint that $L'_{ABCD},
R'_{ABCD}$ be sufficiently suppressed (i.e yielding proton
lifetime $\tau_p > 10^{34}$ yrs)  then flows  invariably towards
parameter regions  where $Z_{H,\ovl H} <<1$ so SO(10) Yukawa
couplings required to match the MSSM are greatly reduced. The same
couplings enter $L'_{ABCD},R'_{ABCD}$ quadratically thus
suppressing them drastically. This mechanism is generically
available to realistic multi-Higgs theories.  Indeed we go so far
as to say that \emph{any} UV completion incapable of performing
this computation will remain less than quantitative and thus is no
viable completion proposal at all. A tedious calculation
determines  the threshold corrections, see \cite{gutupend,
bstabhedge} for details. $Z\simeq 0$ also leads to smaller GUT
couplings and compressed heavy spectra compared to previous fits.
Higher loop  corrections seem computationally prohibitive. However
we have calculated\cite{so10rg,ckphd,ilaPhD} the complete SO(10)
two loop beta functions and   two loop threshold corrections  also
rely upon the same   anomalous dimensions. So convoluting   GUT
scale mass spectra with our SO(10) loop sums determines  two loop
threshold corrections as well. Recovering our one loop results by
this method is  the necessary first step to proceed in this
direction\cite{agcktoappear}.

Searches for fits using the threshold corrected Baryon Decay
operators yield  s-spectra  similar to those found earlier, but
with  smaller values of all couplings (because of the constraints
$Z_{H,\bar H}>0$ imposed on the searches: which  are badly
violated if  almost  any of the super potential parameters grow
large), and acceptable
  $d=5$ B decay rates: as shown in Table 1.  Imposing  $Max|
{O}^{(4)}|< 10^{-22 } GeV^{-1}$   gives proton lifetimes
  above $10^{34}$ yrs.
   $Z_{H,\ovl H}$ approach zero (from above) while $Z_{f,\bar f}$
  are close to 1 : since $\mathbf{16-}$plet Yukawas
 are all suppressed.  See \cite{gutupend,bstabhedge,ckphd,ilaPhD} for further
 details of these and  and many related issues.
\begin{table}
 $$
 \begin{array}{|c|c|c|c|c|c|}
 \hline
 {\rm Solution}&\tau_p(M^+\nu) &\Gamma(p\rightarrow \pi^+\nu) & {\rm BR}( p\rightarrow
 \pi^+\nu_{e,\mu,\tau})&\Gamma(p\rightarrow K^+\nu) &{\rm BR}( p\rightarrow K^+\nu_{e,\mu,\tau})\\ \hline
 1
 &
    9.63 \times 10^{ 34}
 &
    4.32 \times 10^{ -37}
 &
 \{
   1.3 \times 10^{ -3}
 ,
   0.34
 ,
   0.66
 \}&
    9.95 \times 10^{ -36}
 &\{
   4.6 \times 10^{ -4}
 ,
   0.15
 ,
   0.85
 \} \\
 2
 &
    3.52 \times 10^{ 34}
 &
    2.14 \times 10^{ -36}
 &
 \{
   1.7 \times 10^{ -3}
 ,
   0.18
 ,
   0.81
 \}&
    2.62 \times 10^{ -35}
 &\{
   1.8 \times 10^{ -3}
 ,
   0.19
 ,
   0.81
 \} \\
 \hline\end{array}
 $$
\caption{\small{$d=5$ operator mediated proton lifetimes
$\tau_p$(yrs), decay rates
 $ \Gamma ( yr^{-1} )$ and Branching ratios in the dominant Meson${}^++\nu$ channels. }} \label{BDEC}\end{table}

Our  fits are associated with very distinctive sparticle spectra.
An example   is shown in Table 2 which is of the general type used
to evaluate the B-violation rates seen in Table 1. Note the
peculiar and remarkable smallness of the smuon mass which is a
consequence of an RG flow driven by special features of the two
loop  RG flow of soft susy breaking parameters in models with
$M^2_{H}>M^2_{\bar H}<0 $. As discussed in \cite{gutupend} the RG
flow in such theories can be such as to drive $M_{\tilde\tau_R}^2
$ first negative and then to large positive values(due to the
large value of the third generation Yukawa coupling) while in the
relatively flat evolution of the first two s-generations the
   R-smuon lags the  R-stau and the R-selectron the R-smuon due
   to the significant difference in their Yukawa couplings.  This can result
 in the peculiar feature that the running mass squared of the smuon  is
 smallest of all the sfermion masses and close to the LSP mass.
 Then the smuon(or other light sfermion since in other cases up or
 down squarks of the light generations may
  emerge lightest)  provides the necessary co-annihilation channel for
 achieving an acceptable  Bino LSP dark matter \cite{ilaPhD}. The
 smuon case is doubly attractive since it will also contribute to
 the muon g-2 anomaly.  We note however that the sfermion spectra
 shown here are not yet loop corrected and that the loop corrected
 spectra in solutions where we have found so far do not present
 this feature.

\begin{figure}[h]
\includegraphics[width=10.5cm,height=7 cm,clip]{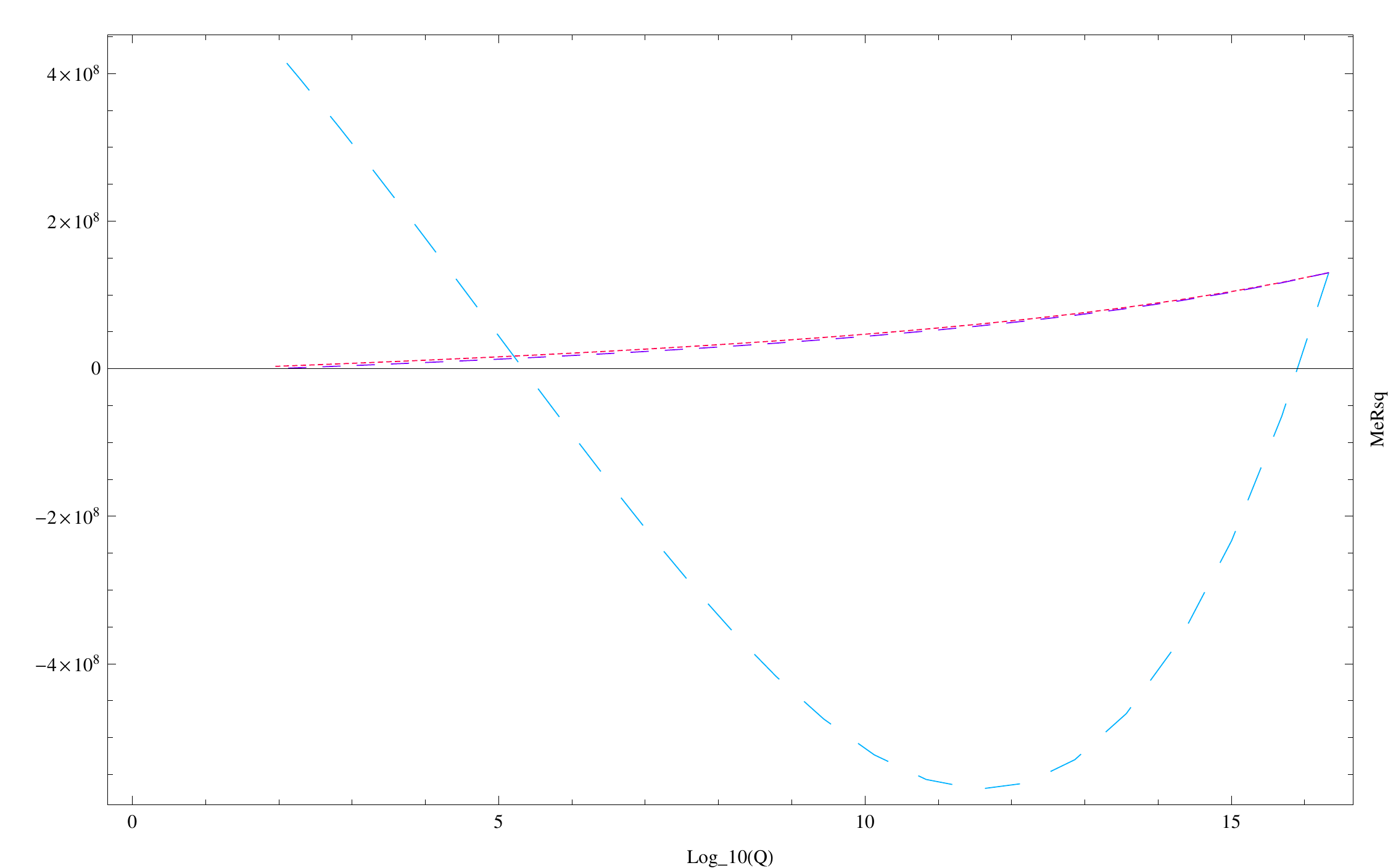}
\caption{Running right-slepton masses squared showing development
of small R-smuon(blue-medium dashed)  as it leads the
R-selectron(small dash-red) and far lags the R-stau(large blue
dashes) which first turns negative and then becomes large driven
by large third generation Yukawa coupling}.
\end{figure}

\begin{table}
 $$
 \begin{array}{|c|c|}
 \hline {\mbox {Field } }&{\rm Mass(GeV)}\\
 \hline
                M_{\tilde{G}}&            1000.14   \\
               M_{\chi^{\pm}}&            569.81,         125591.22\\
       M_{\chi^{0}}&           { {210.10}_{LSP}},            569.81,         125591.20    ,         125591.20\\
              M_{\tilde{\nu}}&         15308.069,         15258.322,         21320.059\\
                M_{\tilde{e}}&           1761.89,          15308.29,           \color{red}  {211.57}_{smuon} \color{black}  ,          15258.60,          20674.72,          21419.56  \\
                M_{\tilde{u}}&          11271.80,          14446.76,          11270.63   ,          14445.80,          24607.51,          40275.87  \\
                M_{\tilde{d}}&           8402.99,          11272.10,           8401.48   ,          11270.95,          40269.19,          51845.93  \\
                        M_{A}&         377025.29\\
                  M_{H^{\pm}}&         377025.30\\
                    M_{H^{0}}&         377025.28\\
                    M_{h^{0}}&         \color{red} {124.00}_{h^0}\color{black} \\
 \hline
 \end{array}
 $$
\caption{\small{Soft Susy breaking spectrum: Large   $\mu,B,A_0 $
  Bino LSP. Light gauginos, Normal S-hierarchy
Higgs ($h^0$) as found , Light smuon ! Other sfermions  multi-TeV
: Decoupled \& Mini-split, large $\mu,A_0$ }}\end{table}

  Besides the immediate satisfaction of finding our
model may be realistic and viable, our calculations underline that
the nature of the competition among candidate UV completions has
been modified   by our results.  Candidate (Susy) UV completion
models with a large or infinite number of fields must not only
show that their parameters can consistently be constrained to
yield some variant of the (MS)SM \emph{but also} that modification
of light wave functions by coupling to heavy fields is both \emph{
calculable} and sensible in the sense of the possibilities
considered and demonstrated above. Absent such a calculability and
consistency,  a UV completion will remain at the level of a
phantasmal possibility rather than a scientific model capable of
braving this falsification gauntlet to progress to the status of a
scientific Theory. We have thus shown that for the NMSGUT careful
attention to the quantum communication between the low energy
effective theory and the UV completion  \emph{through the light
Higgs Portal} yields natural and generic suppression of fast
proton decay in Susy GUTs\cite{bstabhedge} due to the cumulative
effects of the large number of GUT fields that allow the theory to
approach and live on the ``Higgs dissolution edge'' by
renormalizing the light Higgs wavefuntion and thus its Yukawa
couplings, even when individual GUT couplings are all  well within
the perturbative regime.

\section{Supersymmetric Seesaw Inflation} The necessity of  cosmological
Inflation and even its description in terms of of a single
`Inflaton' slowly rolling down a potential plateau before
oscillating around its true minimum to produce quanta of the low
energy theory in the post-inflationary reheating regime is by now
  well accepted\cite{encycinflat}. However the  provenance    of the
Inflaton field and potential is faced by an embarrassing
multiplicity of candidates:   many of which are
compatible\cite{encycinflat} with  even the latest
data\cite{Planck}. From a Particle Physics- rather than
gravitation physics- viewpoint identification of an Inflaton
candidate from among the known particle degrees of freedom is most
appealing. The suggestion \cite{MSSMInflation} to utilize one (or
more) of the D-flat directions(labelled by the Chiral gauge
invariants)  of the MSSM Lagrangian as an inflaton candidate has
thus enjoyed a vogue. Such a suggestion does however face various
stringent constraints such as  ensuring a suitable   scale of
inflation together with the  tiny Yukawa couplings for matter
fermions required to ensure flatness of the potential during
inflation. Since neutrino-lepton Yukawa couplings may well be
small flat directions with sneutrino and lepton components are
well suited for this role. In \cite{akm} an MSSM extended by
conjugate neutrinos($N\equiv\nu^c_L$) superfields to allow (tiny)
\emph{Dirac} neutrino masses and an extra gauge generator(B-L) was
used as a framework for an inflaton made up of the Lepton and
Higgs doublets (the flat direction is labelled as $\phi\equiv
LHN$). Provided a stringent fine tuning is made between the
trilinear soft susy breaking $A{_\phi}$ and the inflaton mass
$\phi$  enjoys a renormalizable potential characterized by an
inflection point at $\phi_0 \sim m_\phi/h_{\nu}$ with $m_\phi \sim
100~ GeV-10 \, TeV, h_\nu\sim 10^{-12}$. Then  $\phi_0\sim 10^{11}
- 10^{15}$ GeV  and  the energy density is characterized by a
scale $V_0^{1/4}\equiv \Lambda\sim m_\phi/\sqrt{h_\nu}\sim :
10^8-10^{10}$ GeV. This model is extremely fine
tuned\cite{lythdimo}
 since the inflaton  mass $M_\phi$ is set by the Susy breaking scale
 the TeV range. Moreover since susy breaking parameters are subject to radiative corrections
 it seems strange to impose extreme fine tuning on the
 trilinear coupling of the inflaton which arises only from the
 soft supersymmetry breaking parameters.
  It was natural\cite{ssI} for
us  to ask the consequences of Seesaw\cite{seesaw} masses for the
light neutrinos i.e. heavy Majorana masses for the right handed
neutrinos and \emph{consequently} tiny Majorana masses for the
left handed neutrinos. In this case we showed that the soft
parameters were irrelevant, tuning was done in the superpotential,
and the large right handed neutrino masses($m_\phi\sim
M_{\nu^c}\sim 10^{6}-10^{12} GeV$) and reasonable superpotential
yukawas $h_{\nu} \sim 10^{-13} (M/GeV)^{1/2}$ which could thus be
as large as $10^{-6}$ and have mild fine tuning: rather than the
radiatively unstable and highly fine tuned\cite{lythdimo} soft
symmetry breaking parameters of the Dirac case. The embedding of
the Supersymmetric Seesaw Inflation(SSI) model in the NMSGUT was
also attempted but could not achieve a large enough number of
e-folds. However the advent of BICEP2 has reopened the whole
question of the inflaton mass and suddenly made it plausible that
it might be anywhere in the range between $M_{\nu^c}- M_{GUT}$ !
The ratio of tensor to scalar modes $r$ emerges as $r\sim
2(m_\phi/(10^{14} GeV))^3 $ but $r$ will be measurable in the near
term  only if $r> 10^{-1.5}$. If the value of $r$ is in  this ball
park then the inflaton mass would need to be much larger than even
the commonly accepted upper limit of around $10^{12} $ GeV on the
right handed neutrino masses. Interestingly the NMSGUT embedding
studied by us earlier allows a generalization where mass of
superheavy Higgs doublets mixed into the  inflaton control its
mass allowing inflaton masses right upto the BICEP2 indicated
range. This scenario allows rasing the achievable number of
e-folds by around 5 orders of magnitude ! The contribution of Ila
Garg in this volume provides further details.

 \section{ Grand Yukawonification and the flavor-Susy breaking link}
The NMSO(10)GUT has achieved  gauge and third generation yukawa
unification consistent with B violation limits, along with
excellent fits of other known fermion data. It makes  and
intriguing and distinctive  predictions regarding sparticle
spectra, and may well be  compatible with the dominant Cold Dark
Matter plus Inflation plus Leptogenesis cosmological scenario
specially if the scale of Inflation turns out to be near that
suggested by BICEP2\cite{bicep2}. It is only natural to ask
whether so complete a  model can pretend to shed any light on the
most outstanding mystery of particle physics : the flavor puzzle
a.k.a the fermion hierarchy i.e. the origin of the peculiar large
inter-generational ratios of charged fermion masses combined with
small quark but two large lepton mixing angles. At first sight,
since we impose no discrete symmetries and since yukawa couplings
are parameters dialed to fit the data, it seems that this is
asking too much. However  reflection on the way in which the SM
Yukawa couplings emerge and the fact that the \emph{only}
convincing hint of flavor unification  seems to be the convergence
of third generation Yukawa couplings at the GUT scale suggests
that perhaps the Yukawa couplings may emerge via spontaneous
breaking of the U(3) flavour symmetry of the SO(10) invariant
matter kinetic terms. In \cite{gyuk} we suggested that our
work\cite{aulmoh,abmsv,ag2,blmdm,nmsgut,bstabhedge} on MSGUTs
 naturally and minimally identifies MSGUT Higgs
multiplets as candidate Yukawon multiplets. It thus yields a novel
mechanism whereby  the fermion hierarchy could emerge from a
flavour symmetric and renormalizable GUT . In  our work
``Yukawons''   \emph{also carry representations of the  gauge
(SM/GUT) dynamics}.  In previous work typically   the  dimension 1
Yukawa-on $\mathcal{Y}$ in  the Higgs vertex makes it
non-renormalizable   (${\cal{L}}= f^c \mathcal{Y} f H
/\Lambda_\mathcal{Y} +...$ : where  the  unknown high scale
$\Lambda_\mathcal{Y}$ controls Yukawa-on dynamics) .
  Minimal SO(10) GUTs\cite{aulmoh,ckn,abmsv,nmsgut}  provide a gauged $O(N_g)$
family symmetry  route to ``Yukawonifcation'' : with the GUT and
family symmetry breaking at the same scale: obviating the need for
non-renormalizable interactions and
 any extraneous scale $\Lambda_\mathcal{Y}$. As we shall see the
  consistency conditions for the maintainability of the   Higgs Portal to the UV completion of the
MSSM play a central role in determining just how the peculiarly
lopsided and ``senseless'' fermion hierarchy  is produced from the
 flavour symmetric and Grand Unified  UV completion.
 In minimal Susy GUTs  one   eschews   invocation  of discrete symmetries
and insists only  upon following the logic of
  SO(10) gauge symmetry. This
insistence,  combined with careful attention to the implications
of the emergence of a single  light  MSSM Higgs pair from the $2
N_g(N_g +1)$ pairs   in the $O(N_g)$ extended MSGUT, leads to an
effectively unique extension of the SO(10) gauge group by a
$O(N_g)$ family gauge symmetry for the $N_g$ generation case and
the dynamical emergence of   fermion hierarchy and mixing. The $
\mathbf{{\oot (\os)}}$ also contributes to both neutrino and
charged  fermion  masses. Gauging just  an  $O(N_g)$
     subgroup of the   $U(N_g) $ symmetry of the fermion kinetic
  terms seems the  workable option : in contrast to a unitary family
  group, because the use of complex representations  introduces
   anomalies and requires doubling of the Higgs
  structure to cancel anomalies and to permit holomorphic  invariants
   to be formed for the superpotential.  Worse,
  Unitary symmetry   enforces vanishing of half the
  emergent matter Yukawa couplings. $O(N_g)$ family symmetry suffers from none
  of these defects and gauging it ensures that no Goldstone bosons
  arise when it is spontaneously broken.
We emphasize  that in contrast with previous
`spurion/Yukawa-on''(see e.g. \cite{koide}) our  model  is
renormalizable and    GUT based.

  The GUT superpotential   has \emph{exactly} the same form as
  the MSGUT (See \cite{ag1,ag2,bmsv,nmsgut} for comprehensive details)  :
  \bea W_{GUT}&=&\mathrm{Tr}( m \Phi^2 + \lambda \Phi^3 + M \Sigb .\s +\eta \Phi .\Sigb.
  \s)\nnu
  &+& \Phi.H.(\gamma \Sigma +\bar\gamma  . \Sigb) + { M_H}  H.H)\label{WGUT}\nnu
  W_{F}&=& \Psi_A .((h H) + (f \s) + (g\Theta) )_{AB}\Psi_B \label{WF}\eea

We have shown how the  ${\bf{120}}$-plet is  included in $W_F$ but
have studied  only  MSGUTs (i.e with $\mathbf{10,\oot}$).
Inclusion of the \textbf{120}-plet does not affect GUT SSB. The
\emph{only} innovation in Higgs structure  is that all
   the MSGUT Higgs fields now carry symmetric
  representation of the $O(N_g)$  family symmetry :
  $\{\Phi,\Sigb,\s,H\}_{AB}; A,B=1,2..N_g$ (under which the
  matter $\mathbf{16}$-plets $\psi_A$ are vector $N_g$-plets).    Couplings $h,f,g$ are
  single complex numbers while the Yukawons
  carry symmetric ($\mathbf{H,\Sigb}$) and anti-symmetric ($\mathbf{\Theta}$)
   representations of   $O(3)$  : as required by the transposition
    property of   relevant  SO(10) invariants. For $N_g=3$,
   real fermion mass parameters come  down from 15
    ($Re[h_{AA}],f_{AB})$   to just
    3 ($Re[h],f$) without the \textbf{120}-plet
 (6 additional to just 2 with the ${\mathbf{120}}$-plet).
  Thus this type of renormalizable flavour unified GUTs can
    legitimately be called Yukawon Ultra-Minimal
    GUTs(\textbf{YUMGUT}s). Further details on the gauge
    symmetry breaking  and determination of Yukawas can be found in the
    contribution of Charanjit Kaur in this volume. We will focus
    on our resolution\cite{bmvacua} of a severe technical
    difficulty that crops up when we spontaneously break a gauged
     O(3) subgroup of the U(3) supersymmetric  flavour symmetry.
     If this difficulty is resolved the determination of the viability of the ``Grand
Yukawonification'' model then becomes a matter of searching the
relatively  small remaining parameter space for viable parameter
sets that fit the fermion data at $M_X$ while taking account of
threshold corrections at low and high scales and while respecting
constraints on crucial quantities like the proton
lifetime\cite{ag2,nmsgut,bstabhedge}. Note that in this approach
not only are the hard parameters of the visible sector
superpotential reduced by replacement of the flavoured parameters
by bland family symmetric ones  but also the soft supersymmetry
breaking parameters are determined by the two parameters of the
hidden sector superpotential and the Planck scale.

    With just the family index carrying MSGUT Higgs present
     the flavor D terms cannot vanish.
     So    additional fields with   vevs  free to cancel the contribution ${\bar{D}_X}^A$
    of the GUT sector to $O(N_g)$ D terms are needed.  The extra  F terms must  be
     sequestered from the GUT sector to preserve the MSGUT SSB.  The special role of
     the Bajc-Melfo supersymmetry breaking is that it provides flat directions in both
the    singlet and the gauge variant parts of a symmetric chiral
supermultiplet $S_{AB}=S_{BA}$.   Since it is very difficult to
make the contribution of the visible sector GUT fields to the
$O(N_g)$ D-terms vanish,   the $\hat S$ flat direction performs
the invaluable function of cancelling this contribution
\emph{without disturbing the symmetry breaking in the visible
sector}.  We proposed \cite{bmvacua}  Bajc-Melfo type
    two field superpotentials \cite{ray,BM} ( of structure $W=S (\mu_B \phi  + \lambda_B \phi ^2)$ ).
    Their   potentials     have   local minima breaking  supersymmetry ($<F_{S}> \neq 0$) which  leave  the vev
    $<S>$ undetermined  ($\phi$ gets a vev). If $\phi,S$ transform in symmetric representations of $O(N_g)$ then
    the undetermined vev of the   trace free part of $S$ is determined at the global supersymmetry level
    by the  minimization of the  $O(N_g)$ D-terms thus  preserving Supersymmetry from high scale breaking. On the other
    hand,   BM superpotentials and their metastable   (local) supersymmetry
    breaking vacua function efficiently\cite{bmvacua} as
    hidden sectors for   supergravity GUT  models :
    coupling to  supergravity determines the $O(N_g)$ singlet part of
    $S$ to have a vev of order the Planck scale. This felicitous
    and unexpected  marriage of the NMSGUT to gauged flavor
    symmetry and Supergravity reduces the number of SO(10) yukawas
    drastically and makes a quite novel linkage between spontaneous violation of  flavor
    and supersymmetry breaking. Since the traceless parts of the  fermionic  components of $S$
    are \emph{massless} at tree level(the trace part is eaten by the
    massive gravitino) they get masses only via  radiative
    corrections and tend to remain very light. They could  thus
    provide very light dark matter candidates as well as cause
cosmological problems of the sort typically associated with moduli
fields\cite{coughlan}. This requires further detailed study  but
it is interesting that such a simple extension of the NMSGUT can
both reduce free parameters and provide moduli dynamics by taking
on the meta-problem of the origin of flavour hierarchy. The
structure used entails yet further stringent constraints since the
masslessness of the   moduli multiplets  $\hat{S}_{AB}$ before
supersymmetry breaking  implies  the existence of $N_g(N_g+1)/2
-1$ SM singlet fermions generically lighter than the gravitino
mass scale and possibly as light as a few GeV. In addition the
Polonyi mode $S_s$ may  also lead  to difficulties in the
cosmological scenario. Thus such modes can be both a boon and a
curse for familion GUT models. A boon because generic Susy GUT
models are hard put if asked to provide Susy WIMPs of mass   below
100 GeV as CDM candidates as suggested by the DAMA/LIBRA
experiment\cite{damalibra}. A curse because there are strong
constraints on the existence of such light moduli which normally
demand that their mass be rather large ($> 10$ TeV)
 due to the robust   cosmological(`Polonyi') problems
 arising from   decoupled modes with Planck scale VEVs\cite{modulicon}.
  In contrast to the simple Polonyi model and
  String moduli, the BM moduli  have explicit   couplings  to light fields through
family D-term mixing and loops. Moreover the MSGUT scenario
favours\cite{nmsgut,bstabhedge}  large gravitino masses $> 5-50 $
TeV. Thus  the Polonyi  and moduli problems may be evaded.  In any
case the cosmology need be considered seriously only   after we
have shown that the MSSM fermion spectrum is indeed generated by
the ``Yukawonified" NMSGUT\cite{yukawon}.

 \section{Discussion }

The vast range of applicability of the
 NMSGUT well fits it to claim that it is a theory for all epochs.
 Not only does it  present a well controlled framework for
 realizing the long standing dream of Grand Unification in a
 completely realistic fashion, it also makes distinctive
 predictions concerning the observation of supersymmetry and
 lepton flavour violation and dynamically palliates the proton
  decay problem  that has chronically afflicted Susy GUTs without introducing
 extraneous structures but in a way intrinsic to the GUT itself.
 Furthermore it provides a surprisingly simple context for the
 embedding of high scale supersymmetric inflation. The right
 handed neutrinos and adequate- CKM-PMNS linked-  CP violation that
  are present can enable Leptogenesis. Reaching still
 further it even serves as a basis for the generation of the
 flavour hierarchy by the spontaneous violation of a gauged
 flavour symmetry at the GUT scale and in the process makes the
 startling suggestion that supersymmetry breaking is linked to
 flavour violation via a novel  BM hidden sector. This  implies
 a striking and novel(for Susy GUTs) prediction  of possibly very light
 non-neutralino  Dark Matter candidates. Interestingly the self
 coupling of this light DM is not constrained. The working out of the
various flavour breaking  scenarios while maintaining realism in
the fermion and
 GUT SSB sector and accounting for low and high scale threshold
 corrections is indeed a formidable but    manageable
 project that has already  cleared several
 hurdles\cite{gmblm,blmdm,nmsgut,bstabhedge} that might well have
 falsified this model in the three decades since its proposal.
 These successes motivate our belief that a focus on the
 implication of the consistency conditions that define our world
 as we know it in the context of the  flavour symmetric NMSGUT may
 further extend the scope of the NMSGUT to include an  dynamical
 understanding of the flavour hierarchy. Moreover the strong reduction in
 the number of parameters makes falsification much more
 practicable.  Although it continues to live dangerously, we   hope that this
 will be seen as a scientific virtue  of the NMSGUT and its generalizations
  rather  than as an unnecessary
 concreteness and optimism:  with which it is sometimes reproached.
\bibliographystyle{pramana}

\bibliography{references}

\end{document}